\documentclass{article}

\usepackage{amsmath,epsfig,amssymb}
\usepackage{graphicx}
\usepackage{eufrak,mathrsfs}
\usepackage[mac]{inputenc}
\usepackage{array} 
\usepackage{upgreek,bm} 
\usepackage{lettrine}

\def\0{\boldsymbol{0}}
\def\1{\boldsymbol{1}}
\def\x{\bm{x}}

\def\0{\boldsymbol{0}}
\def\1{\boldsymbol{1}}
\def\bw{\boldsymbol{\omega}}
\def\u{\boldsymbol u}

\def\k{{\boldsymbol k}}

\def\I{\mathscr I}

\def\H{\mathpzc H}
\def\Hil{\mathscr H}

\newcommand{\mbb}{\mathbb}

\newcommand{\w}{\omega}

\DeclareSymbolFont{EUr}{U}{eur}{m}{n}
\DeclareSymbolFont{EUb}{U}{eur}{b}{n}
\DeclareMathSymbol{\varphi}{\mathord}{EUr}{"27}
\DeclareMathAlphabet{\mathpzc}{OT1}{pzc}{m}{it}

\newtheorem{theorem}{Theorem}[section]

\newtheorem{proposition}[theorem]{Proposition}

\newenvironment{proof}[1][Proof.]{\begin{trivlist}
\item[\hskip \labelsep {\bfseries #1}]}{\end{trivlist}}

\title{Gabor wavelet analysis and the fractional Hilbert transform} 

\author{Kunal Narayan Chaudhury and  Michael Unser \\
Biomedical Imaging Group, \\ 
Ecole Polytechnique Fédérale de Lausanne (EPFL),  Switzerland
}

\date{}

\begin{document} 
\maketitle


\begin{abstract}

We propose an amplitude-phase representation of the dual-tree complex wavelet transform (DT-$\mbb{C}$WT) which provides an intuitive interpretation of the associated complex wavelet coefficients. The representation, in particular, is based on the \textit{shifting} action of the group of fractional Hilbert transforms (fHT) which allow us to extend the notion of arbitrary phase-shifts beyond pure sinusoids. We explicitly characterize this shifting action for a particular family of Gabor-like wavelets which, in effect, links the corresponding dual-tree transform with the framework of windowed-Fourier analysis. 

	We then extend these ideas to the bivariate DT-$\mbb{C}$WT based on certain directional extensions of the fHT. In particular, we derive a signal representation involving the superposition of direction-selective wavelets affected with appropriate phase-shifts.	
\end{abstract}

\section{INTRODUCTION}

\subsection{The dual-tree transform}

\lettrine{W}{e} begin by briefly reviewing the fundamentals of the dual-tree transform. The transform involves a pair of  wavelet bases with a one-to-one `quadrature' correspondence between the basis elements \cite{kingsbury2,selesnick}. Specifically, one considers a \textit{primary} wavelets basis $\{\psi_{i,k}\}_{(i,k) \in \mathbf{Z}^2}$ of $\mathrm{L}^2(\mathbf{R})$ generated through the dilation-translations of a single prototype $\psi(x)$; that is, $\psi_{i,k}(x)=\Xi_{i,k} \psi(x)$ where $\Xi_{i,k} f(x)=2^{i/2} f(2^i x-k)$ denotes the (normalized) dilation-translation operator corresponding to integers $i$ and $k$. The highlight of the transform is then the construction of a \textit{secondary} wavelet basis $\{\psi'_{i,k}\}_{(i,k) \in \mathbf{Z}^2}$ having the correspondence $\psi_{i,k}'(x)=\Hil \psi_{i,k}(x)$, where $\Hil$ denotes the Hilbert transform (HT) operator: 
\begin{equation}
\label{HT_def}
\Hil f(x) \stackrel{\mathscr{F}}{\longleftrightarrow} -j \ \mathrm{sign}(\w)\hat{f}(\w).
\end{equation}
The HT acts as a quadrature transform that takes $\cos(\w_0x)$ into $\sin(\w_0x)$, and as an orthogonal transform on $\mathrm{L}^2(\mathbf{R})$ in the sense that $\langle f, \Hil f \rangle=0$ for all $f(x)$ in this space. Though this is not at all obvious a priori, it turns out (as suggested by the notation) that the secondary wavelet basis can also be realized through the dilations-transations of the HT counterpart $\psi'(x)=\Hil \psi(x)$. This, in fact, is possible thanks to certain fundamental invariances enjoyed by the HT operator. In particular, following definition \eqref{HT_def}, one can readily verify that the HT commutes with translations and dilations; in particular,
\begin{equation}
\label{inv1}
\Hil \ \Xi_{i,k}=\Xi_{i,k} \Hil;
\end{equation}
and that it is unitary:
\begin{equation}
\label{inv2}
||\Hil f||_{ \mathrm{L}^2}=||f||_{\mathrm{L}^2} \qquad (f \in \mathrm{L}^2(\mathbf{R})).
\end{equation}
It is then easily deduced that the functions $\psi'_{i,k}(x)=\Xi_{i,k} \psi'(x)$ indeed constitute a wavelet basis of $\mathrm{L}^2(\mathbf{R})$, and that the correspondence $\psi'_{i,k}(x)=\Hil\psi_{i,k}(x)$ holds for every integer $i$ and $k$ \cite{kunal_journal}.

The application of the transform involves the simultaneous analysis of a signal $f(x)$ in $\mathrm{L}^2(\mathbf{R})$ in terms of  the quadrature wavelet bases $\{\psi_{i,k}\}$ and $\{\psi'_{i,k}\}$. In particular, one considers the wavelet expansions
\begin{align}
\label{branches}
f(x)= \begin{cases}  \sum_{(i,k) \in \mathbf{Z}^2} a_i[k] \psi_{i,k}(x),  \\
\sum_{(i,k) \in \mathbf{Z}^2} b_i[k] \psi'_{i,k}(x),
\end{cases}
\end{align}
where the expansion coefficients in \eqref{branches} are specified by the dual wavelet bases $\{\tilde \psi_{i,k}\}$ and $\{\tilde \psi'_{i,k}\}$ through the projections 
\begin{equation}
\label{analysis_comps}
a_i[k]=\langle f, \tilde \psi_{i,k} \rangle, \quad  \mbox{and}  \quad b_i[k]=\langle f, \tilde  \psi'_{i,k} \rangle.
\end{equation}
In effect, this allows one to identify the complex wavelet coefficients $c_i[k]=(a_i[k]+j b_i[k])/2$, and the associated amplitude-phase factors $|c_i[k]|$ and $\mathrm{arg} (c_i[k])$ (the use of the factor $1/2$ will be justified shortly). As a consequence of \eqref{inv1} and \eqref{inv2}, the dual wavelet bases can also be generated through the dilations-translations of two dual wavelets, $\tilde \psi(x)$ and $\tilde \psi'(x)$, that form a HT pair as well.

\subsection{Multiresolution Gabor-like transforms}

	A framework for constructing HT-pairs of wavelets (within Mallat's multiresolution framework) was recently proposed based on a spectral factorization result for scaling functions \cite{kunal_journal}. In particular, it was shown that a multiresolution form of Gabor-like analysis could be achieved within the framework of the dual-tree transform. This was founded on the two vital observations. The first one was that the extended $(\alpha,\tau)$ family of B-spline wavelets $\psi(x;\alpha)$ (indexed by the approximation order $\alpha+1$) is closed with respect to the action of the HT operator (the corresponding discrete wavelet transform has an efficient FFT-based implementation). Secondly, it was shown that the complex spline wavelet $\Psi(x)=\psi(x; \alpha)+j \Hil \psi(x; \alpha)$ asymptotically converges to a Gabor function:
\begin{equation*}
\Psi(x;\alpha) \sim \varphi(x) \ \mathrm{exp}\big(j\w_0 x+\xi_0\big) \qquad (\alpha \rightarrow +\infty),
\end{equation*} 
where $\varphi(x)$ is a Gaussian window, and $\w_0, \xi_0$ are appropriate modulation parameters. As a consequence, a Gabor-like transform, involving the computation of the sequence of projections
\begin{equation}
\label{WFT}
f(x) \mapsto \frac{1}{2} \left\langle f(x), \Xi_{i,k} \Psi(x;\alpha) \right \rangle \qquad (i,k \in \mathbf{Z})
\end{equation}
with the dilates-translates of the Gabor-like wavelet $\Psi(x;\alpha)$, could be realized using the dual-tree transform corresponding to the spline wavelets $\psi(x;\alpha)$ and $\Hil \psi(x;\alpha)$  (sufficiently large $\alpha$). These ideas were also extended for the realization of a direction-selective Gabor-like transform.

\subsection{Present Contribution}
	
	In this paper, we provide a characterization of the dual-tree transform, and the Gabor-like transforms in particular, from the perspective of multiresolution windowed-Fourier analysis. In particular, we link the multiresolution wavelet-framework of the former with the intuitive amplitude-phase representation associated with the latter. Complex wavelets, derived from the combination of non-redundant wavelet bases, provide an attractive means of encoding the relative signal ``displacements'' using the phase relation between the components. The DT-$\mbb{C}$WT is a particular instance where the components are related through the HT. 
	
	In \S\ref{one-dimensional-case}, we derive a representation of the dual-tree transform using the group of fractional Hilbert transform (fHT) operators:
\begin{equation}
\label{def_fHT}
\H_{\tau}=\cos(\pi\tau)\ \I - \sin(\pi\tau) \ \Hil \qquad (\tau \in \mathbf{R})
\end{equation}
($\I$ is the identity operator). In particular, we are able to interpret the phase factors associated with the dual-tree transform in terms of the action of this group. However,  it is the fundamental invariances \eqref{inv1} and \eqref{inv2} inherited by this extended family of operators that play a decisive role in establishing the windowed-Fourier-like representation (cf.  \eqref{WFA}). 
	       
	The proposed windowed Fourier-like representation admits a straightforward extension to the bivariate setting by introducing an appropriate multi-dimensional extensions of the HT. In particular, we arrive at a representation (cf. \eqref{2D_amp_phase}) involving the superposition of the direction-selective synthesis wavelets affected with appropriate phase-shifts. This provides an explicit understanding of the phase-shift action of the fdHT operators for a the particular family of 2D Gabor-like wavelets derived through the tensor products of $1$D Gabor-like wavelets.

\section{DUAL-TREE GABOR WAVELET ANALYSIS}
\label{one-dimensional-case}

 \subsection{Signal representation: interpretation of the amplitude-phase factors}
 \label{representation}
 
 Our objective is to derive a representation of $f(x)$ in terms of the amplitude-phase factors $c_i[k]=|c_i[k]|\mathrm{e}^{j\phi_i[k]}$. Clearly, the transformation
\begin{equation*}
f(x) \mapsto \big\{c_i[k]\big\}_{(i,k) \in \mathbf{Z}^2}
\end{equation*}
constitutes a overcomplete representation of $f(x)$. In particular, given the coefficients $c_i[k]$, there exists non-unique ways of reconstructing the input $f(x)$. We consider the simplest inversion procedure involving the inversion of both the forward transforms, as in \eqref{branches}, followed by the averaging of the reconstructed signals. In particular, by combining the expansions in \eqref{branches} and by invoking the dilation-translation invariance of the fHTs, we arrive at the following representation:
\begin{align}
\label{amp-phase}
f(x)&= \frac{1}{2} \sum_{(i,k) \in \mathbf{Z}^2} \Big(a_i[k] \psi_{i,k}(x)+b_i[k] \psi'_{i,k}(x)\Big)   \nonumber \\
& =  \sum_{(i,k) \in \mathbf{Z}^2} |c_i[k]|  \H_{\phi_i[k]/\pi} \big\{\psi_{i,k}(x)\big\}  \nonumber \\
& = \sum_{(i,k) \in \mathbf{Z}^2} |c_i[k]| \ \Xi_{i,k} \big\{\psi(x; \tau_i[k])\big\}.
\end{align}
Here the synthesis wavelet $\psi(x;\tau_i[k])$ is derived from the mother wavelet $\psi(x)$ through the action the fHT corresponding to the shift $\tau_i[k]=\phi_i[k]/\pi$. The unitary nature of the fHT ensures that these fractionally-shifted synthesis wavelets have identical norms. In particular, while the amplitude $|c_i[k]|$ indicates the strength of wavelet correlation, the local signal displacement gets encoded in the shift $\tau_i[k]$ which specifies the most ``appropriate'' wavelet within the family $\{\H_{\tau} \psi_{i,k}\}_{\tau \in \mathbf{R}}$.

\subsection{Characterization of the Gabor-like transform}

	It turns out that the shifted wavelets $\psi(x;\tau_i[k])$ in \eqref{amp-phase} can be explicitly characterized when $\psi(x)$ is a (real) Gabor-wavelet,
\begin{equation}
\label{real_gabor_wavelet}
\psi(x)=\varphi(x) \cos\big(\w_0 x + \xi_0\big).
\end{equation}
This formula describes the asymptotic form of the Gabor-like wavelet $\Psi(x;\alpha)$ in \eqref{WFT}. For reasons that will be evident shortly, we choose to flip the roles of the analysis and synthesis wavelets: we will analyze the signal using the dual complex wavelet $\tilde \Psi(x;\alpha)=\tilde \psi(x;\alpha)+j \tilde \psi'(x;\alpha)$, while the Gabor-like wavelet $\Psi(x;\alpha)$ will be used for reconstruction.

	If the window function $\varphi(x)$ is bandlimited to $[-\Omega,\Omega]$ with $\Omega <\w_0$, we can make precise statements on the dual-tree representation in \eqref{amp-phase}. To do so, we will need the following result:
			
\begin{proposition} 	
\label{main_result}	
	 Let $\upvarphi(x)$ in \eqref{real_gabor_wavelet} be bandlimited to $(-\w_0,\w_0)$. Then
\begin{equation}
\label{fHT_action}
\H_{\tau} \big \{ \upvarphi(x) \cos(\w_0x)\big \}=\upvarphi(x) \cos(\w_0x+\pi \tau).
\end{equation}
\end{proposition} 
That is, the fHT acts on the phase of the modulating sinusoid while preserving the Gaussian envelope. In particular, we can then rewrite \eqref{amp-phase} as
\begin{equation}
\label{WFA}
f(x) = \sum_{(i,k) \in \mathbf{Z}^2} \stackrel{\mathrm{fixed \ window}}{\overbrace{\upvarphi_{i,k}(x)}} \ \Xi_{i,k}  \Big\{ \stackrel{\mathrm{variable \ amp-phase \ oscillation}}{\overbrace{ \big|c_i[k]\big| \cos\big(\w_0 x+\xi_0+\pi \tau_i[k]\big)}}\Big\}
\end{equation}
where $\upvarphi_{i,k}(x)=\Xi_{i,k} \upvarphi(x)$ denotes the (fixed) Gaussian-like window at scale $i$ and translation $k$. This provides an explicit interpretation of the parameter $\tau_i[k]$ as the phase-shift applied to the modulating sinusoid of the wavelet. In effect, the oscillation is shifted to best fit the underlying signal singularities/transitions while the localization window $\upvarphi_{i,k}(x)$ is kept fixed. In this light, one can interpret the associated dual-tree analysis as a multiresolution form of the windowed-Fourier analysis, with the fundamental difference that, instead of analyzing the signal at different frequencies, it resolves the signal over different scales (or resolutions).

\begin{figure*}
\centering
\fbox{\includegraphics[width=0.8\linewidth]{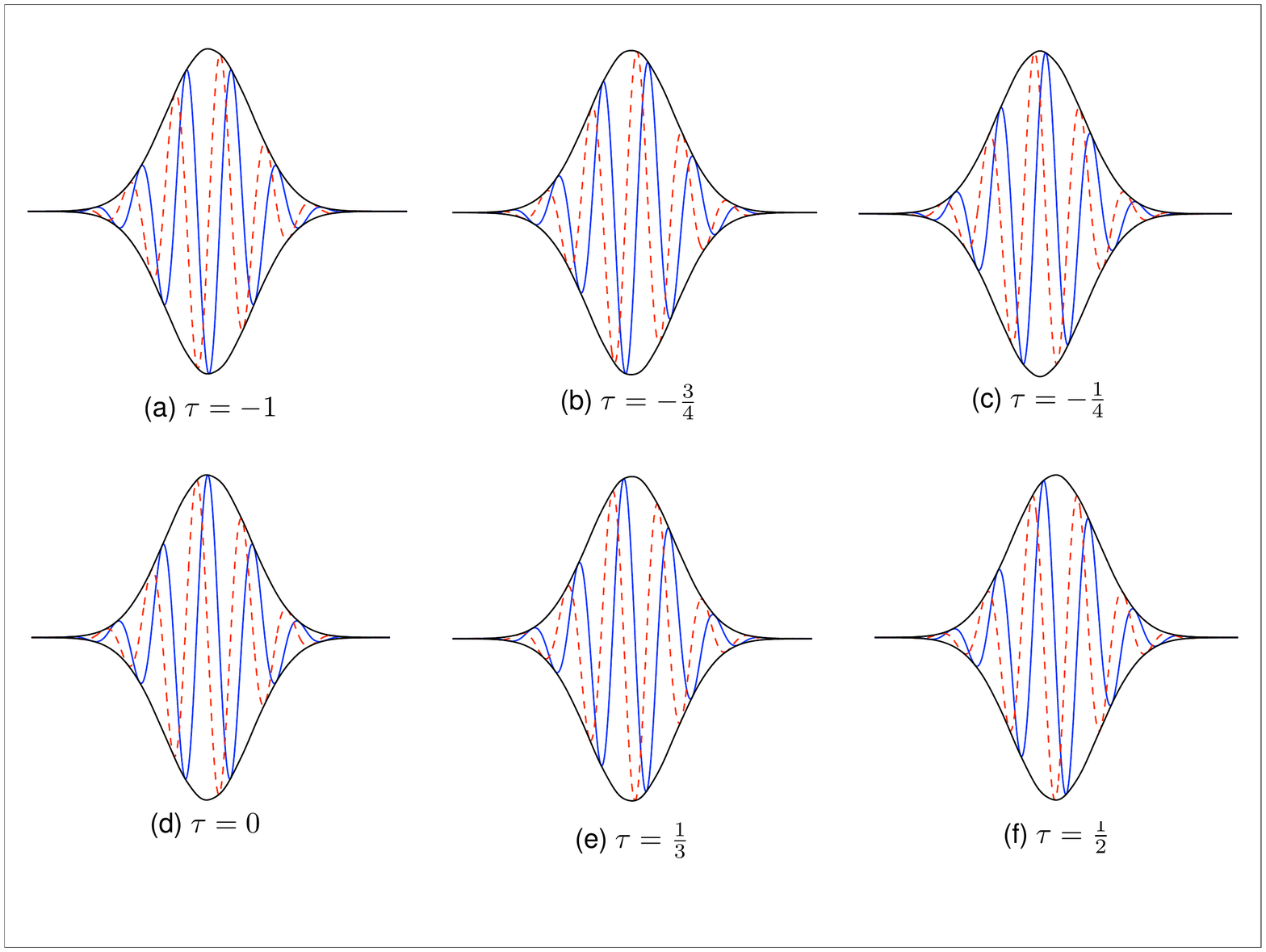}} 
\medskip
\caption{Quadrature pairs of Gabor-like spline wavelets obtained by the action of fHT group. Blue (solid line): $\H_{\tau} \psi(x;8)$, Red (broken line): $\H_{\tau+\frac{1}{2}} \psi(x;8)$, and Black (solid line): Common localization window given by  $|\H_{\tau} \psi(x;8) +j\H_{\tau+\frac{1}{2}}  \psi(x;8)|$ (http://dx.doi.org/doi.number.goes.here).}
\label{fig}
\end{figure*}

	Figure \ref{fig} shows quadrature pairs $(\H_{\tau} \psi(x;\alpha),\H_{\tau+1/2}  \psi(x;\alpha))$ of Gabor-like spline wavelets corresponding to different $\tau$. Each of the pairs are localized within a common Gaussian-like window, and the modulating oscillations are driven to a relative quadrature through the action of the pair $(\H_{\tau},\H_{\tau+\1/2})$.

\section{BIVARIATE EXTENSION}
\label{directional_wavelets}

		The amplitude-phase representation derived in \S \ref{one-dimensional-case} can also be extended to the $2$D setting where the dual-tree wavelets exhibit better directional selectivity than the conventional tensor-product (separable) wavelets \cite{CTDWT}.

\subsection{Directional HT pairs of wavelets}

	We briefly recall the construction framework for the bivariate DT-$\mbb{C}$WT based on the tensor-products of one-dimensional analytic wavelets  \cite{kunal_journal}. Specifically, let $\varphi(x)$ and $\varphi'(x)$ denote the scaling functions associated with the analytic wavelet $\psi_a(x)=\psi(x)+j\psi'(x)$, where $\psi'(x)=\Hil \psi(x)$. The $2$D dual-tree construction then hinges on the identification of four separable multiresolutions of $\mathrm{L}^2(\mathbf{R}^2)$ that are naturally associated with the two scaling functions: the approximation subspaces $V(\varphi)\otimes V(\varphi), V(\varphi) \otimes V(\varphi'), V(\varphi') \otimes V(\varphi)$ and $V(\varphi') \otimes V(\varphi')$, and their multiscale counterparts. The corresponding separable wavelets -- the `low-high', `high-low' and `high-high' wavelets -- are specified by:
\begin{alignat}{2}
\label{separable_wavelets}
\bar{\psi}_1(\x)&=\varphi(x)\psi(y), &  \hspace{8mm}  \bar{\psi}_{4}(\x)&=\varphi(x)\psi'(y),  \nonumber \\
\bar{\psi}_{2}(\x)&=\psi(x)\varphi(y),  & \hspace{8mm}  \bar{\psi}_{5}(\x)&=\psi(x)\varphi'(y),  \nonumber \\
\bar{\psi}_{3}(\x)&=\psi(x)\psi(y), &  \hspace{8mm}  \bar{\psi}_{6}(\x)&=\psi(x)\psi'(y),  \nonumber \\  \nonumber \\
\bar{\psi}_{7}(\x)&=\varphi'(x)\psi(y) , & \hspace{8mm}  \bar{\psi}_{10}(\x)&=\varphi'(x)\psi'(y), \nonumber \\ 
\bar{\psi}_{8}(\x)&=\psi'(x)\varphi(y), & \hspace{8mm} \bar{\psi}_{11}(\x)&=\psi'(x)\varphi'(y), \nonumber \\ 
\bar{\psi}_{9}(\x)&=\psi'(x)\psi(y), & \hspace{8mm}  \bar{\psi}_{12}(\x)&=\psi'(x)\psi'(y).
\end{alignat}
The dual wavelets $\tilde{\bar{\psi}}_1(\x),\ldots,\tilde{\bar{\psi}}_{12}(\x)$ are similarly defined in terms of $\tilde \psi(x)$ and $\tilde \psi'(x)$ (here $\x=(x,y)$ denotes the planar coordinates). As far as the identification of the complex wavelets is concerned, the main issue is the poor directional selectivity of the `high-high' wavelets along the diagonal directions. This problem can, however, be mitigated by appropriately exploiting the one-sided spectrum of the analytic wavelet $\psi_a(x)$, and, in effect, by appropriately combining the wavelets in \eqref{separable_wavelets}. In particular, the complex wavelets specified by
\begin{align}
\label{def_CW}
\Psi_1(\x) &= \psi_a(x) \varphi(y) \ =\bar{\psi}_2(\x)+j\bar{\psi}_8(\x) ,    \nonumber \\
\Psi_2(\x) &= \psi_a(x)  \varphi'(y)=\bar{\psi}_5(\x)+j \bar{\psi}_{11}(\x),   \nonumber \\
\Psi_3(\x) &= \varphi(x) \psi_a(y) \ =\bar{\psi}_{1}(\x)+j\bar{\psi}_{4}(\x),     \nonumber \\
\Psi_4(\x) &= \varphi'(x) \psi_a(y)=\bar{\psi}_{7}(\x)+j\bar{\psi}_{10}(\x),   \nonumber \\
\Psi_5(\x) &= \frac{1}{\sqrt{2}}\psi_a(x)  \psi_a(y)= \left(\frac{\bar{\psi}_{3}(\x)-\bar{\psi}_{12}(\x)}{\sqrt{2}}\right)+j\left(\frac{\bar{\psi}_{6}(\x)+\bar{\psi}_{9}(\x)}{\sqrt{2}}\right), \nonumber \\
\Psi_6(\x) &= \frac{1}{\sqrt{2}}\psi^{\ast}_a(x)  \psi_a(y) =\left(\frac{\bar{\psi}_{3}(\x)+\bar{\psi}_{12}(\x)}{\sqrt{2}}\right)+j\left(\frac{\bar{\psi}_{6}(\x)-\bar{\psi}_{9}(\x)}{\sqrt{2}}\right),
\end{align}
exhibit the desired directional selectivity along the primal orientations $\theta_1=\theta_2=0$, $\theta_3=\theta_4=\pi/2$, $\theta_5=\pi/4$, and $\theta_6=3\pi/4$, respectively \cite{kunal_journal}. The dual complex wavelets $\tilde \Psi_1(\x),\ldots,\tilde \Psi_6(\x)$ are specified in an identical fashion using the dual wavelets $\tilde{\bar{\psi}}_{p}(\x)$, and are oriented along the same set of directions. \newline

	Akin to the HT correspondence, the complex wavelet components are related through the directional HT (dHT):
\begin{equation}
\label{directionalHT}
\Hil_{\theta} f(\x) \stackrel{\mathscr{F}}{\longleftrightarrow} -j\mathrm{sign}(\u_{\theta}^T \bw)\hat{f}(\bw) \qquad (0 \leqslant \theta < \pi),
\end{equation}
where $\u_{\theta}=(\cos \theta,\sin \theta)$ denotes the unit vector along the direction $\theta$. In particular, we have the correspondences 
\begin{equation*}
\mathfrak{Im} (\Psi_{\ell})= \Hil_{\theta_{\ell}} \mathfrak{Re}(\Psi_{\ell}) \quad (\ell=1,\ldots,6),
\end{equation*}
so that, by denoting the real component of the complex wavelet $\Psi_{\ell}(\x)$ by $\psi_{\ell}(\x)$, we have the convenient representation $\Psi_{\ell}(\x)=\psi_{\ell}(\x)+j \Hil_{\theta_{\ell}} \psi_{\ell}(\x)$ that is reminiscent of the $1$D analytic representation. 

\subsection{Directional amplitude-phase representation}

	Let us denote the dilated-translated copies of the each of the six analysis wavelets $\tilde\Psi_{\ell}(\x)$ by $\tilde \Psi_{\ell,i,\k}(\x)$, so that
\begin{equation*}
\tilde \Psi_{\ell,i,\k}(\x)=\Xi_{i,\k} \tilde\Psi_{\ell}(\x) \quad  (i \in \mathbf{Z}, \k \in \mathbf{Z}^2),
\end{equation*}
where $\Xi_{i,\k}$ is specified by $\Xi_{i,\k}f(\x)=2^i f(2^i \x-\k)$. The corresponding dual-tree transform involves the analysis of a finite-energy signal $f(\x)$ in terms of the sequence of projections
\begin{equation*}
c^{\ell}_i[\k]= \frac{1}{4} \big \langle f, \tilde \Psi_{\ell, i, \k} \big \rangle.
\end{equation*}
The representation of $f(\x)$ in terms of the analysis coefficients $c^{\ell}_i[\k]$ is based on the following fractional extension of the directional HT operator:
\begin{equation*}
\H_{\theta, \tau}=\cos(\pi\tau)\ \I - \sin(\pi\tau) \ \Hil_{\theta} \qquad (\tau \in \mathbf{R}).
\end{equation*}
These operators allow us to capture the notion of direction-selective phase-shifts. The key properties of the fHT, which played a decisive role in establishing the representation for the $1$D counterpart, carry over directly to the fractional directional HT (fdHT) operators: they are invariant to translations and dilations and are unitary. In particular, based on the above properties, we derive the representation
\begin{equation}
\label{2D_amp_phase}
f(\x)= \sum_{(\ell,i,\k)} \big |c^{\ell}_i[\k] \big|  \ \Xi_{i,\k} \big\{\psi_{\ell} \big(\x ; \tau^{\ell}_i[\k]\big)\big\}
\end{equation}
involving the superposition of direction-selective synthesis wavelets affected with appropriate phase-shifts \cite{submission}. The wavelets $\psi_{\ell}\big(\x ; \tau^{\ell}_i[\k]\big)$ are derived from the reference wavelet $\psi_{\ell}(\x)$ through the action of fdHT, corresponding to the direction $\theta_{\ell}$ and shift $\tau^{\ell}_i[\k]=\arg(c^{\ell}_i[\k])/\pi$ As in the $1$D setting, further insight into the above representation is  obtained by considering wavelets resembling windowed plane waves.

\subsection{Directional Gabor-like analysis}

	Akin to the $1$D setting, further insight into the above representation is  obtained by considering wavelets resembling windowed plane waves. A distinctive feature of the dHT, that comes as a direct consequence of \eqref{directionalHT}, is its phase-shift action in relation to plane-waves: it transforms the directional cosine $\cos(\u_{\theta}^T\x)$ into the directional sine $\sin(\u_{\theta}^T\x)$. Moreover, what turns out to be even more crucial in the current context, is that the above action is preserved for windowed plane waves of the form
\begin{equation}
\label{plane_wave}
\varphi(\x) \cos(\Omega \u_{\theta}^T\x).
\end{equation}
In particular, as a straightforward directional extension of \eqref{fHT_action}, we have the following generalization for the fractional extensions:

\begin{proposition} Suppose that $\varphi(\x)$ in \eqref{plane_wave} is bandlimited to the disk $\{\bw: ||\bw|| < \Omega\}$. Then we have that
\begin{equation}
\label{modulation_fdHT}
\H_{\theta,\tau} \left\{ \varphi(\x) \cos(\Omega \u_{\theta}^T\x) \right\}=\varphi(\x) \sin(\Omega \u_{\theta}^T\x+\pi \tau).
\end{equation}
\end{proposition}
Thus, the fdHT acts only on the phase of the oscillation while the window remains fixed. In particular, if the dual-tree wavelets are of the form $\psi_{\ell}(\x) = \upvarphi_{\ell}(\x) \cos\left(\Omega_{\ell} \u_{\theta_{\ell}}^T\x\right)$, we can then rewrite \eqref{2D_amp_phase} as  
\begin{equation}
\label{WFA2}
f(\x) = \sum_{(\ell,i,\k)} \stackrel{\mathrm{fixed \ window}}{\overbrace{\upvarphi_{\ell, i,\k}(\x)}} \ \Xi_{i,\k}  \Big\{\stackrel{\mathrm{variable \ amp-phase \ directional \ wave}}{ \overbrace{\big |c^{\ell}_i[\k]\big| \cos\left(\Omega_{\ell} \u_{\theta_{\ell}}^T\x+\pi \tau^{\ell}_i[\k]\right)}} \Big\}, 
\end{equation}
where $\upvarphi_{\ell, i,\k}(\x)$ represent the dilated-translated copies of $\varphi_{\ell}(\x)$. The above representation explicitly  highlights the role of $\tau^{\ell}_i[\k]$ as a ``scale-dependent'' measure of the local signal displacements along certain preferential directions. This is the scenario for the spline-based Gabor-like transforms \cite{kunal_journal} where the dual-tree wavelets asymptotically converge to directional Gabor functions. 

\section{CONCLUDING REMARKS}

We presented an amplitude-phase representation of the dual-tree transform in general, and a windowed-Fouier-like characterization of the Gabor-like transforms in particular. The signal representation was centered around one crucial construction, namely the HT correspondence between the wavelet bases. Indeed, the identification of the fHT-transformed wavelets in \eqref{WFA} followed as a direct consequence of this particular relation; the subsequent developments were then based on two crucial properties of the fHT, namely   
\begin{itemize}
\item its intrinsic invariances with respect to translations, dilations and norm-evaluations, and
\item its particular phase-shifting action on the Gabor wavelet. 
\end{itemize}
These observations could be of potential interest in applications involving the dual-tree transform, particularly signal denoising, where a rigorous mathematical model linking the reconstructed signal to the processed complex wavelet coefficients is desirable.  

\section*{Appendix: Proof of Proposition \ref{main_result}}

\begin{proof}
The result follows directly from definition \eqref{def_fHT} and the following action of the HT: 
\begin{equation}
\label{HT_action}
\Hil \left\{ \varphi(x) \cos(\w_0 x) \right\} = \varphi(x) \sin(\w_0 x).
\end{equation}
Indeed, we see that
\begin{align*}
\H_{\tau} \big \{ \upvarphi(x) \cos(\w_0x)\big \}&=\cos(\pi \tau)\ \varphi(x) \cos(\w_0 x) - \sin(\pi \tau) \ \Hil \left\{ \varphi(x) \cos(\w_0 x) \right\} \\
&=\cos(\pi \tau)\ \varphi(x) \cos(\w_0 x) - \sin(\pi \tau) \ \varphi(x) \sin(\w_0 x) \\
&= \upvarphi(x) \cos(\w_0x+\pi \tau).
\end{align*}
To establish \eqref{HT_action}, we note that the Fourier transform\footnote{we use $\hat{f}(\w)=\int_{\mathbf R} f(x) \exp{(-j \w x)} \mathrm{d} x$ as the definition of the Fourier transform of $f(x)$.} of $\varphi(x) \cos(\w_0 x)$ is given by $\pi (\hat\varphi(\w-\w_0)+\hat\varphi(\w+ \w_0))$. Following definition \eqref{HT_def}, we then have that 
\begin{align*}
\Hil \left\{ \varphi(x) \cos(\w_0 x) \right\} \stackrel{\mathscr{F}}{\longleftrightarrow} &  -j \ \mathrm{sign}(\w)\cdot \pi \big(\hat\varphi(\w-\w_0)+\hat\varphi(\w+\w_0)\big) \\
&  =- j \pi \big(\hat\varphi(\w-\w_0)-\hat\varphi(\w+\w_0)\big) \\
\stackrel{\mathscr{F}}{\longleftrightarrow} & \ \varphi(x) \sin(\w_0 x),
\end{align*}
since $- j\pi (\hat\varphi(\w-\w_0)-\hat\varphi(\w+\w_0))$ is the Fourier transform of $\varphi(x) \sin(\w_0 x)$. Note that, in going from the first to the second step, we have used the crucial fact that the supports of $\varphi(\w+\w_0)$ and $\varphi(\w-\w_0)$ are entirely restricted to the half-lines $\{\w <0\}$ and $\{\w>0\}$, respectively.
\end{proof}

\bibliography{bibliograph}   
\bibliographystyle{spiebib}   

\end{document}